# Multilayer Effects in $Bi_2Sr_2Ca_2Cu_3O_{10+z}$ Superconductors

Giulio Vincini*, Setsuko Tajima*, Shigeki Miyasaka* and Kiyohisa Tanaka**

*) Dept. of Physics, Osaka University, Osaka 560-0043, Japan

**) Institute of Molecular Science, Okazaki 444-8585, Japan

Abstract

We discuss the characteristic features of triple Cu-$O_2$ layer cuprates superconductors, by comparing those of single and double layer cuprates superconductors. After a brief introduction to multilayer cuprates and their characteristic properties such as the doping imbalance between the inner and outer Cu-$O_2$ planes (IP and OP, respectively) revealed by nuclear magnetic resonance, we present the experimental results of angle resolved photoemission and Raman scattering spectroscopy for the triple layer $Bi_2Sr_2Ca_2Cu_3O_{10+z}$ which showed two different superconducting gaps opening on the IP and OP. The doping dependence of the double peak structure in Raman spectra was found to be qualitatively consistent with that of single and double layer cuprates, if each layer doping for the IP and OP is taken into account. The fact that the IP and OP share the same electronic phase diagram and the same transition temperature ($T_c$) hints to a coupling between the IP and OP. Moreover, the energies of IP and OP Raman peaks were found to be very large, not scaling with $T_c$, which can be attributed to the strong influence of the pseudogap of the underdoped IP in triple layer cuprates. These findings suggest that the high $T_c$ and the large gap ratio of triple layer cuprates are realized through a combination of the interlayer coupling between the OP and IP and the interaction between superconductivity and the pseudogap.

E-mail: tajima@phys.sci.osaka-u.ac.jp

# 1. Introduction

More than thirty years have passed since the discovery of high temperature superconductivity in the cuprate[1]. This discovery has brought us many interesting research subjects; the mechanisms of high temperature superconductivity and many other emergent phenomena in strongly correlated system such as charge confinement[2-4], spin/charge striped order[5], electronic liquid crystal[6], pseudogap[7], precursor of superconductivity[8-11]. However, despite considerable research efforts, the mechanisms of these emergent phenomena have not been elucidated.

In this review, we focus on the multilayer effect on superconducting gap size. It is well known that cuprate superconductors have a layered structure and the superconducting transition temperature $T_c$ changes with the number of layers in a unit cell[12]. The origin of this $T_c$ variation is still unclear. One idea is that the multilayer structure protects the inner $CuO_2$-plane from a buckling distortion that may suppress the $T_c$ value[13,14]. Another model is that the multilayer structure plays an active role to enhance $T_c$ [15-17].

Because the cuprates superconductors have a strong electron correlation and two dimensional crystal structure, the carriers are confined to the $CuO_2$-layers in the normal state, which can be proved by the incoherent charge transport along the $c$-axis (out-of-plane direction)[18]. In other words, we cannot adopt a so-called band picture which gives bonding-, antibonding and non-bonding bands for triple layer compounds.

The superconducting state is also two dimensional, because the superconducting coherence length is shorter than the interlayer distance[19,20]. The system can be regarded as an alternative stack of superconducting and insulating layers. In such a circumstance, it is a very interesting question whether each $CuO_2$-layer is completely independent or not. A nuclear magnetic resonance (NMR) experiment indicated that the carrier doping levels are different between the inner- and outer-$CuO_2$ layers[21]. Then, each layer could have different $T_c$ values. Nevertheless, we observe only one bulk transition. The question is, if there is interlayer coupling, what determines the bulk $T_c$ value and superconducting gap size?

To answer these questions, we studied the superconducting state in $Bi_2Sr_2Ca_2Cu_3O_z$, a typical triple layer cuprate. As the superconducting gap in the cuprates has a $d$-wave symmetry, we need $k$-selective experimental probes. Raman scattering spectroscopy (RSS) and angle-resolved photoemission spectroscopy (ARPES) are the most powerful $k$-selective tools. As the gap sizes estimated from these two techniques are not always identical, we have examined a direct comparison of RSS and ARPES through the Kubo formula analysis. Both probes successfully detected the two superconducting gaps that are inherently part of each of outer and inner CuO$_2$ planes. Based on the gap sizes and their doping dependences, the possibility of interlayer coupling and its effect on superconductivity will be discussed.

**2. Charge imbalance revealed by NMR**

The physical properties of a series of multilayer cuprates have been intensively studied by Mukuda and his collaborators[21] through NMR, which is a site selective probe. Figure 1 illustrates the crystal structures of a typical family of cuprates with different numbers of CuO$_2$-layers, n, in a unit cell. In a single layer compound (n=1), the layers of Cu-O octahedra are sandwiched by HgO blocking layers. A double layer compound (n=2) contains two Cu-O pyramids separated by a Ca-layer. In a triple layer compound (n=3), the Cu-O square layer is inserted between the two pyramids. When n becomes larger than 3, the number of inserted Cu-O square layers increases. While there is only one (equivalent) Cu-site in the case of n=1 and 2, the Cu-sites in the Cu-O square and the pyramid are inequivalent for n=3 and 4. In this way, the number of inequivalent sites increases with increasing n. Here we name the CuO$_2$-layer within a pyramid an "outer plane (OP)", while calling the Cu-O square layer an "inner plane (IP)".

The advantage of NMR is that it can distinguish the electronic state on each atomic site; thus, information about the inner- and outer-CuO$_2$ plane can be resolved. A method for estimating the carrier doping level of each layer from the Knight shift data has been established. Figure 2 demonstrates a linear correlation between the spin part of the Knight shift for B//ab ($K_s^{ab}$) and the hole doping level $p$ estimated from the $T_c$ value[22, 23] for the compounds with n=1 and 2 where all the Cu-sites are equivalent[21]. Then, this relation was applied to estimate the hole doping level of each layer from $K_s^{ab}$ even for compounds with inequivalent Cu-sites.

A clear double peak was observed in $^{63}$Cu-NMR spectra for Bi$_2$Sr$_2$Ca$_2$Cu$_3$O$_{10+\delta}$ (Bi2223) and other multi-layer cuprates with n≥3. (An example is shown in Fig.3[24]) This indicates a difference in $K_s^{ab}$ for the IP and OP, namely, a difference in doping level between the IP and OP. A broader peak is expected to correspond to the OP, because the effect of disorders such as a bucking distortion must be stronger in the OP near the blocking layer. As a result, it was revealed that the doping level is always lower in the IP than in the OP. This is likely if we consider the fact that the carriers are supplied from the blocking layers. However, the inhomogeneous charge distribution in a metal is unusual. As we see in the incoherent c-axis charge dynamics, it is not appropriate to treat this system as a conventional metal. We need to consider each CuO$_2$-layer independent or electronically separated. Along the c-direction, it may be better to regard this system as an ionic compound. This means that the system is really two dimensional (2D) and is not like other strongly anisotropic three dimensional materials such as organic superconductors [25].

### 3. $T_c$ and doping control of Bi2223 crystals

The unusual metallic picture mentioned above for multi-layer cuprates has been confirmed by ARPES and Raman scattering measurements for Bi2223, as will be described in the next section. Before discussing these data, it is worth to note some properties of Bi2223 crystals.

Single crystals of Bi2223 were grown by a traveling solvent floating zone method[26]. The carrier doping level is controlled by changing the oxygen content through post-annealing under several conditions. The variation of oxygen content can be monitored by the $c$-axis lattice parameter, as demonstrated in Fig.4[27]. As the oxygen content increases, the $c$-axis lattice parameter decreases. Therefore, the $c$-axis lattice parameter is a measure of hole carrier concentration. A similar tendency is observed also in Bi2212. The peculiarity of Bi2223 is that $T_c$ is almost constant in the high doping region, which is different from the general behavior of many other cuprates such as Bi2212. The authors speculated that this is due to the redistribution of holes between the IP and OP that occurs with the incorporation of a large number of oxygen. It suggests that the carrier doping levels of IP and OP are not identical.

Transport properties such as electrical resistivity in the $ab$- and $c$-direction and thermopower were precisely measured[27]. Figure 5 shows the temperature dependence of resistivity both in the $ab$-plane and the $c$-direction for various oxygen concentrations. With increasing oxygen concentration, resistivity in both directions systematically decreases. This indicates that carrier doping monotonically proceeds with increasing oxygen concentration.

In contrast to this monotonic change of resistivity, $T_c$ variation is not monotonic with oxygen concentration. It is well known that the cuprates show a characteristic doping dependence of $T_c$. Namely, with increasing the carrier doping level ($p$) $T_c$ first increases and after reaching a maximum value it turns to decrease, forming a dome-like $T_c(p)$ curve. We call the doping level with a maximum $T_c$ "optimum doping", while the lower and higher doping than the optimum are called as under- and over-doping, respectively.

The ARPES measurements were carried out on the optimally doped Bi2223 that were grown by Uchida's group at the University of Tokyo[28]. (The detailed information about their crystals is not available.) The crystals for our Raman study were grown by Watanabe's group at Hirosaki University, and the carrier concentration was changed by ourselves at Osaka University. The crystals with four different doping levels were prepared. The magnetic susceptibility of these samples is shown in Fig.6[29]. The $T_c$ value of optimally doped crystal was $T_c$ =109K, while those of the two underdoped crystals were 105K and 88K. We prepared one overdoped crystal with $T_c$=109K. Hereafter, we refer to these samples as OpD109 (optimally doped with $T_c$=109K), UnD105 (underdoped with $T_c$=105K), UnD88 (underdoped with $T_c$ =88K), and OvD109 (overdoped with $T_c$ =109K), respectively. All the sample-related information are summarized in Table I.

Although the $T_c$ values of OpD109 and OvD109 are almost the same, we confirmed that carrier doping level of OvD109 is higher than in the optimally doped sample (OpD109), by measuring the $c$-axis lattice parameters. As shown in Table I, the $c$-axis of OvD109 is shorter than that of OpD109, indicating that the oxygen content is higher in the former sample.

If the carrier doping levels are different between the IP and OP, it is unclear which layer determines the bulk $T_c$. Here, we assume that the average doping $p_{av}$ of the two layers determines $T_c$ with an empirical relation[22, 23] as follows.

$$T_c/T_{c,max} = 1 - 1.82(p_{av} - 0.16)^2 \quad (1)$$

(Here the doping level is defined by the deviation of average Cu valence from two. The doping level of 0.16 gives a maximum $T_c$.) Using this relation, we can estimate $p_{av}$ from the $T_c$ value. The average doping levels for UnD105 and UnD88 were estimated as $p_{av}$=0.14 and 0.11, respectively. For the OvD109 sample, however, we cannot use this method to estimate $p_{av}$. Then, as the first approximation for OvD109, $p_{av}$ was estimated from its $c$-axis lattice parameter ($c$) by assuming a linear correlation between $c$ and $p_{av}$. As shown in Fig.7, by extrapolating the data for the other three doping samples, $p_{av}$ of OvD109 was estimated as 0.17.

### 4. Observation of two gaps in optimally doped Bi2223
### 4.1 ARPES observation of two bands and two gaps

The unusual metallic picture for multi-layer cuprates, namely, the doping level difference in IP and OP has been confirmed by ARPES measurements. Figures 8(a) and (b) show the intensity mapping of ARPES spectra of optimally doped Bi2223 near the Fermi energy (E=±20meV)[28]. Two Fermi surfaces (FSs) are clearly resolved. The authors attributed the FS closer to the Γ point to that of the OP and the other to that of the IP, assuming that the doping level is higher in the OP than in the IP. The ARPES intensity of FS strongly depends on the incident photon energy owing to the matrix element effect. This effect is also seen in the band dispersions in the nodal direction (Fig.8(c) and (d)). The photon energy of 7.65 eV gives a stronger intensity in the OP than in the IP, while the energy of 11.95 eV gives an opposite result.

The observed band splitting in Bi2223 is apparently similar to that in Bi2212 [30]. However, it is worth noting that the origin of band splitting is different in these two cases. In Bi2212, two $CuO_2$-planes in a unit cell are equivalent and thus only one FS is observed in the optimally and under-doped samples. The band splitting called bilayer splitting becomes visible only in heavily overdoped samples where the electronic state is closer to the Fermi liquid state and the many body effect is weakened.

By contrast, the two FSs observed in Bi2223 are ascribed to the two inequivalent electronic states of the IP and OP. There are a couple of reasons for this interpretation. First, the crystal is almost optimally doped (not overdoped). Second, the momentum dependence of the superconducting gap is different in these two FSs. Third, we cannot see a non-bonding band that should appear together with bonding and antibonding bands for triple layer compounds when the many body effect is weakened.

Figure 9 plots the extracted gap values against the *d*-wave momentum function. For the OP, the gap energy follows the *d*-wave function, while the gap of the IP largely deviates from the *d*-wave line in the anti-nodal region. Such gap behavior (deviation from the *d*-wave function) is commonly observed in the underdoped samples of mono- and double layer cuprates[31-33]. In other words, a strong enhancement of the energy gap towards (π,0) is considered as evidence that this band is for the underdoped state.

**4.2 Raman observation of two gaps**

Following the pioneering work of ARPES, our research group has studied the Raman scattering spectra of optimally doped Bi2223 crystals. Here, we focus on spectra with $B_{1g}$ and $B_{2g}$ symmetries. The former spectra are obtained with the x'y' cross polarization (*x'*- and *y'* are the directions of incident and scattered light polarization, respectively), while the latter with xy polarization. Here, the *x* and *y* axes are along the Cu-O bond direction in the tetragonal notation, and the *x'* (*y'*) direction makes an angle of 45 degree from the *x*(*y*) directions. Although both polarization spectra contain the $A_{2g}$ component, its intensity is so weak that we can ignore it.

The incident and scattered light polarizations determine the spectral intensity weight through the Raman vertex. As demonstrated in Fig.10, the $B_{1g}$ spectrum mainly probes the electronic state near (±π, 0) and (0, ±π) in the *k*-space, while the $B_{2g}$ mainly probes that near (±π /2, ±π /2). In the case of cuprates, the $B_{1g}$ spectrum sufficiently reflects the anti-nodal region of the Fermi surface (FS) where the $d$-wave gap has a maximum, while the $B_{2g}$ reflects the FS near the gap node.

Figure 10 shows the $B_{1g}$ and $B_{2g}$ Raman scattering spectra at 10K for the OpD109 sample. In the $B_{1g}$ spectra (Fig.11a), in addition to several phonon peaks such as the 255 cm$^{-1}$ peak, there is a contribution from the electronic Raman scattering that monotonically increases with energy from $\omega=0$ and saturates at higher energies. As the temperature is lowered, two changes are observed. One is a suppression of spectral intensity below 700-800cm$^{-1}$, which can be ascribed to the pseudogap opening in the anti-nodal region. Although we cannot precisely determine the pseudogap temperature from this data, it is clearly higher than 115K, which is consistent with the resistivity result for the optimally doped Bi2223[27].

The other change in the $B_{1g}$ spectra with temperature is the appearance of two peaks at around 560 cm$^{-1}$ and 800 cm$^{-1}$. Although the lower energy peak partially overlaps with one of the phonon peaks, a broad and intense peak around 560 cm$^{-1}$ as well as the other peak at 800 cm$^{-1}$ can be ascribed to a superconducting pair-breaking peak, because both peaks appear below $T_c$,. A double pair-breaking peak is the evidence for the presence of two superconducting gaps in this system. This is in sharp contrast with the single gap in single and double layer cuprates such as Bi2212.

Following the ARPES and the NMR results, we assigned the lower energy peak to the pair-breaking peak for the OP and the higher energy one to that for the IP. It is commonly observed that the $B_{1g}$ gap energy decreases with increasing $p$ (doping level) when $p>0.11$[34, 35]. Therefore, the present assignment of the two peaks is consistent with previous Raman results.

Here it is interesting whether these two gaps open simultaneously or not. Although the NMR measurement suggested that the transition temperatures for the IP and OP are different[21], the resistivity and magnetic susceptibility indicate only one step transition at $T_c$, as shown in Figs.5 and 6[36]. In our Raman measurement, as the peaks gradually grew, we could not find any appreciable difference in the critical temperatures of these two peaks.

Next, we discuss the Raman scattering with $B_{2g}$ polarization (Fig.11b). In the $B_{2g}$ spectrum of Bi2223, there is no clear signature of the pseudogap opening and a very broad peak is developed below $T_c$. The peak energy is lower than either of the two $B_{1g}$ peak energies. All these spectral features are similar to the well-known $B_{2g}$ behavior in Bi2212. As a double pair-breaking peak was observed in the $B_{1g}$ spectrum, we expected to observe a double peak also in the $B_{2g}$ spectrum. However, the $B_{2g}$ spectrum exhibited only a single broad peak. This is probably because, as the gap energies for the IP and OP near the node are close to each other, we cannot resolve the two gap peaks. A broad spectral shape of the $B_{2g}$ pair-breaking peak also makes it difficult to distinguish the two gaps in the $B_{2g}$ spectrum.

**4.3 Comparison of the Raman and ARPES results**

To confirm our interpretation of the double peak found in the $B_{1g}$ Raman spectrum of Bi2223, we calculated the Raman spectra from the ARPES, using the Kubo formula. For the calculation of the electronic Raman scattering spectrum, the kinetic theory successfully explained the Raman response as a *d*-wave superconductor[37]. However, it cannot sufficiently reproduce a real spectral shape because the density of states along the Fermi surface is treated as a delta function. To obtain more realistic spectra, we have developed a new method using the Kubo formula. In a previous study on the double layer Bi2212 [38, 39], we proved that this method is valid and advantageous. For triple layer compounds, by separating the IP and OP bands of ARPES, we can calculate their separate contribution to the Raman spectra, and verify if the two $B_{1g}$ Raman peaks truly originate from the two separate bands.

The Kubo formula [34, 40] gives the Raman susceptibility $\chi''_{\gamma\Gamma}$ as:

$$\chi''_{\gamma\Gamma} = \frac{2}{\pi V} \sum_{k} \gamma_k \Gamma_k \cdot \int_{-\infty}^{\infty} (f_\omega - f_{\omega+\Omega}) \, G''_{k,\omega} G''_{k,\omega+\Omega} \left(1 - \frac{\Delta_k^2}{(\omega+\xi_k)(\omega+\Omega+\xi_k)}\right) d\omega \quad (2)$$

Here $f_\omega$ is the Fermi Dirac function, $G''_{k,\omega}$ is the Green function, $\Delta_k$ is the superconducting gap, $\xi_k$ is the normal state band and $\gamma_k$ and $\Gamma_k$ are the bare and renormalized Raman vertex, respectively. The Green function can be obtained from the ARPES intensity $I_{k,\omega}$, by:

$$I_{k,\omega} = I_0 \cdot M_k \cdot f_\omega \cdot A_{k,\omega} \quad (3)$$
$$G''_{k,\omega} = -\pi A_{k,\omega} \quad (4)$$

Here we assumed that the matrix element $M_k$ is constant. As the matrix element has only a small momentum dependence, this assumption is valid as a first approximation.

The ARPES data were obtained by Ideta and his coworkers on a slightly underdoped (but almost optimally doped) sample with $T_c$=108K grown by the Uchida group in the University of Tokyo[28]. The experiment was performed at UVSOR facility BL7 at the National Institute for Molecular Science in Japan. The incident photon energy was $h\nu$=8eV with S polarization and the energy resolution was $\Delta E$=7meV. The sample temperature was $T$=12K. From these spectra we subtracted the background of the inelastically scattered electrons using the Shirley background formula[38,39]:

$$bg_{Shirley}(\omega) = c \int_{\omega}^{\infty} P(\omega') \, d\omega' \quad (5)$$

The ARPES spectra were symmetrized across the Fermi level to obtain the unoccupied part of the spectral function. (This procedure is allowed in the superconducting state.) The normal state band was obtained from a tight binding fit using the equation,

$$\xi_{\boldsymbol{k}} = -2t(\cos k_x a + \cos k_y a) + 4t' \cos k_x a \cos k_y a - 2t''(\cos 2k_x a + \cos 2k_y a) - \mu \quad (6)$$

Here, to reduce the number of free parameters, the next-next nearest hopping integral $t''$ was fixed as half of the value of the next nearest hopping integral $t'$. The Fermi vectors on the ARPES cuts were used as fitting points.

As a first approximation, we considered the renormalized Raman vertex to be equal to the bare one, for which we used the equation calculated under the assumption of single band given by the tight binding formula in a tetragonal lattice:

$$\gamma_{B_{1g},\boldsymbol{k}} = \boldsymbol{\Gamma}_{B_{1g},\boldsymbol{k}} = ma^2 t(\cos k_x a - \cos k_y a) \quad (7)$$

$$\gamma_{B_{2g},\boldsymbol{k}} = \boldsymbol{\Gamma}_{B_{2g},\boldsymbol{k}} = 4ma^2 t' \sin k_x a \sin k_y a \quad (8)$$

The ARPES intensities for the IP and OP bands were separated by a Gaussian fit of the Energy Distribution Curves (EDCs) using three Gaussian peaks, one for the IP and OP band each and one for the high energy incoherent intensity that originates from the strong correlations effects in the antinodal part of the momentum space. (See Fig.12.)

Two attributes of the ARPES dataset make the calculation challenging: (i) For the incident photon energy of 8eV, the OP band intensity is much stronger than the IP band intensity owing to the ARPES matrix element effect, which makes the IP band extremely weak in the nodal and antinodal regions and (ii) the ARPES dataset does not cover the most antinodal part of the momentum space. The first issue forced us to fix some of the fitting parameters, such as peak position and width, in the most nodal and antinodal ARPES cuts where the IP band intensity was very weak. Unfortunately, owing to the strong oscillation of the relative intensity of the two bands with varying incident photon energy[28], there are few datasets available that clearly show both bands in the entire Brillouin zone. The second issue may slightly affect the peak position as a small portion of the antinodal momentum space will be left out of the calculation.

The intensities of the IP and OP bands were normalized to the same value across the entire Fermi surface. It is not clear what the correct intensity ratio of the two bands should be for calculation. One consideration is that, as there are two OPs in a unit cell, the OP band should be more intense than the IP one. However, many factors possibly affect the band intensity of ARPES. More calculation details are described in Ref. [41].

The calculated $B_{1g}$ and $B_{2g}$ Raman spectra for the optimally doped Bi2223 are compared with the experimental ones in Fig.13. One can find that the calculated spectra from the ARPES data successfully reproduce the experimental Raman spectra. The striking result is that, in the $B_{1g}$ (antinodal) configuration, the IP and OP bands exhibit peaks at different energies that are close to the experimental $B_{1g}$ peak energies. The calculated and experimental OP peaks sit at 500 cm$^{-1}$ and 565 cm$^{-1}$, respectively, while the calculated and experimental IP peaks sit at 782 cm$^{-1}$ and 805 cm$^{-1}$, respectively. Here the OP peak position is slightly underestimated, which may be due to the fact that a small portion of the antinodal part of the momentum space is missing in our input ARRPES data. This problem does not affect the IP peak as the IP ARPES band intensity is strongly suppressed in the antinode, namely, the data-missing *k*-region.

By summing the separate contribution of IP and OP, we obtain a thick orange line. A rather good correspondence of the calculated and experimental IP and OP peaks provides strong proof that the double $B_{1g}$ Raman peak truly originates from the two separate bands of Bi2223 and, therefore, that it is a signature of the double superconducting gap of this material.

Also for the $B_{2g}$ Raman spectrum, both of the calculated IP and OP contributions roughly reproduce the broad experimental Raman spectrum, as seen in Fig.13(b). Here the IP and OP band contributions are similar. This is reasonable, considering the small difference between the two SC gaps in the nodal region as well as the broadness of the $B_{2g}$ peak. Although the IP contribution shows three humps, this is caused by the poor fitting quality of the extremely weak IP band in the nodal cut of ARPES and thus is an artifact. The total sum of the IP and OP contribution represents well the overall experimental spectrum.

## 5. Doping dependence of the two gaps in Raman scattering spectra of Bi2223

The $B_{1g}$ and $B_{2g}$ Raman spectra of the OpD109 sample were sufficiently reproduced from the ARPES data, as demonstrated in the previous section. The reproduced spectra justify our assignment of two $B_{1g}$ peaks to the pair-breaking peaks for the IP and OP.

Next, we determined the doping dependence of the Raman scattering spectrum of Bi2223. Figure 14 presents the $B_{1g}$ and $B_{2g}$ Raman spectra of the OvD109, OpD109, UnD105, and UnD88 samples[29]. For the $B_{1g}$ spectra, the low energy suppression due to opening of the pseudogap is commonly observed in all four samples, while the double pair-breaking peak can be seen only in the OvD109, OpD109 and UnD105 samples. In the UnD88 spectra, no clear peak manifests itself and only the pseudogap opening is observed, which is a common behavior in cuprates with $p<0.11$.[34,35] Both of the two peak energies for the IP and OP shift with doping. By contrast, in the $B_{2g}$ spectra, only a single broad peak is observed at the lowest temperature. The peak slightly changes when the carrier doping changes. Here there is no low energy suppression due to the opening of the pseudogap, which is consistent with the previous reports on Bi2212[42,43].

To demonstrate the shift in peak energy more clearly, we calculated the difference spectra $I(10K) - I(T > T_c)$, which indicate the temperature evolution of Raman intensity below $T_c$. (see Fig.15) In the $B_{1g}$ spectra, one can see the systematic shift of both peaks towards higher energies with reducing hole concentration. In addition, we found a weak but clear peak structure also in the UnD88 sample, although no peak was resolved in the raw spectrum shown in Fig.14. It is natural to assign this peak to the pair-breaking peak for the OP because the IP is more underdoped than the OP and thus a pair-breaking peak could not be seen. Actually the intensity of the higher energy peak for OP gradually decreases with underdoping.

In contrast to that of the $B_{1g}$ spectra, the shift of the $B_{2g}$ peak is not drastic. The main spectral change with doping is broadening of the peak, while the peak energy also slightly shifts with doping. As mentioned above, we assume that the peak energy in the $B_{2g}$ spectrum represents the average of the pair-breaking energies for the IP and OP in the nodal direction.

To plot the doping dependence of the gap energies, we need to estimate the doping levels ($p$) for the IP and OP. For the optimally doped Bi2223, $p$ for the IP and OP were estimated from the NMR measurement[24] as $p$(IP)=0.127 and $p$(OP)=0.203, respectively. As there are no available data for the other doping samples, we assume that the charge imbalance between IP and OP is constant with doping. Namely, $p$(OP)−$p$(IP) =0.076 for all the samples. Then, we can determine $p$(IP) and $p$(OP) from the shift of $p_{av}$ from the optimum value ($p$=0.16). The estimated $p$(IP) and $p$(OP) are listed in Table I. Despite such a rough approximation, these values seem reasonable. For example, it is known that the pair-breaking peak disappears in the $B_{1g}$ spectrum when $p$ becomes smaller than 0.11[34,35]. The absence of $B_{1g}$ pair-breaking peak only in the IP ($p$=0.079) of the UnD88 sample is consistent with this general doping dependence of the $B_{1g}$ spectrum.

Using these $p$-values, we plotted the pair-breaking peak energy (PE) as a function of $p$ (Fig.16). There are three $B_{1g}$ data points for the IP and four points for the OP. The $B_{1g}$ pair-breaking energy decreases almost linearly with increasing $p$. The striking result is that all the data points fall on a single line, which indicates that the IP and OP share the same PE($p$) line. This observation also means that a double pair-breaking peak is caused merely by a difference in $p$ in the IP and OP but not by any other factor.

For the $B_{2g}$ pair-breaking peak energy, we could not plot the data of each layer because only a broad single peak was observed in all the samples. Therefore, here, the $B_{2g}$ peak energy is plotted as a function of $p_{av}$. It is unclear whether or not the doping dependence of $B_{2g}$ peak energy follows the $T_c$ dome, whereas such a behavior is observed in many other cuprates[42, 44]. First, the observed peak position is possibly affected by the overlapped two peaks for the IP and OP. Second, the error bars of peak energies are very large because the $B_{2g}$ peak profiles are so broad that there is an uncertainty in determining the peak energy. Finally, the covering range of doping in this study is limited near the optimal doping, namely, it is not wide enough to conclude the doping dependence of the $B_{2g}$ peak energy in Bi2223. If we could achieve more overdoping and/or underdoping, a clearer conclusion would be drawn.

## 6. Discussion

The doping dependences of the $B_{1g}$ and $B_{2g}$ pair-breaking peak energies in Bi2223 (Fig.14) are qualitatively consistent with the results of single- and double-layer cuprates, if the doping levels of the IP and OP are independently taken into account. The monotonic increase in the $B_{1g}$ peak energy does not imply an increase in the $d$-wave gap maximum with decreasing $p$. As reported in the ARPES on Bi2212 and Bi2223, when the pseudogap is opening, the $k$-dependence of the gap is deviated from a simple $d$-wave form and enhanced in the antinodal $k$-region. This is considered as an effect of the pseudogap. Here the antinodal gap does not contribute to the superconductivity condensate, although the gap energy is enhanced. The present Raman data also indicates a strong enhancement of the antinodal gap energy owing to the pseudogap. Note that no antinodal gap feature is observed in the heavily underdoped regime with p<0.11 because of the loss of coherent quasiparticles in the antinodal region. The effect of the pseudogap also manifests itself in the $B_{2g}$ spectrum through the shrink of Fermi arc where the superconducting gap opens. The apparent doping dependence of the $B_{2g}$ gap should not be considered as an indication that the $d$-wave gap maximum follows the $T_c$ dome. It is primarily determined by the Fermi arc length[45,46].

The difference between Bi2223 and the single- and double-layer cuprates is seen in the gap energy scale. Here we introduce another scale, the normalized peak energy PE/$k_B T_{c,max}$, which is shown on the right axis of Fig.14, where $T_{c,max}$ is the maximum $T_c$ value in each compound. This should be 4.2 in a $d$-wave BCS superconductor in a weak coupling limit. However, in the case of Bi2223, it varies from ~7 to ~11 for the $B_{1g}$ peak, while it is about 5 for $B_{2g}$. These values are much larger than those reported for the single- and double-layer cuprates in which most of the data points collapse into universal curves[42], as is presented by the dashed line and curve in Fig.14.

Here it would be interesting to refer to the report of the Raman spectra for another triple layer cuprate, $HgBa_2Sr_2Cu_3O_z$ (Hg1223)[47]. In the $B_{1g}$ Raman spectra of Hg1223, a double pair-breaking peak is also visible, although the authors do not consider it an intrinsic feature. Plotting the PE/$T_{c,max}$ of Hg1223 in Fig.14, we found that the data point sits on the same $B_{1g}$ line. Therefore, we conclude that this large pair-breaking energy scale is a common property of triple layer cuprates.

With regard to the origin of high $T_c$ values for triple layer cuprates, there have been several proposals. According to the tunneling mechanism of Cooper pairs between the $CuO_2$ planes, $T_c$ increases with the number of layers[15,16], although the optical spectroscopy experiment indicated that this effect is negligible[48]. A simpler explanation is that the IP is protected from the out-of-plane disorder and this enhances $T_c$ and the gap value[13,14]. In this scenario, the large gap in the disordered OP is a result of the proximity effect from the IP. Another explanation is that $T_c$ increases with $-t'/t$, where $t'$ and $t$ are the next-nearest-neighbor and the nearest-neighbor hopping integral, respectively[49]. As $-t'/t$ increases with the number of layers, $T_c$ of triple layer cuprates is higher than that of single- and double-layer cuprates. The inter-layer coupling was more positively considered by Kivelson[17]. According to his model, the charge imbalance between the IP and OP plays a key role in enhancing $T_c$. In other words, a higher $T_c$ in the multilayer cuprates can be achieved by combining the underdoped IP with a large pairing energy (represented as a pseudogap energy) but a low superfluid density and the overdoped OP with a small gap energy but a high superfluid density.

Here, we emphasize that the most important finding in Fig.16 is not a large energy gap but a large gap ratio (PE/$T_{c,max}$) in triple layer compounds. Even though we consider a large energy gap and/or a high $T_c$ value owing to several reasons mentioned above, we cannot explain this large ratio. Then, we speculate that the original $T_c$ may be much higher but the actual bulk $T_c$ is suppressed for an as yet unidentified reason. It is plausible that the suppression might be due to the proximity effect of the IP with a lower $T_c$.

Regarding the interlayer coupling, we note that there is only one superconductivity transition in Bi2223, as we observed in resistivity and magnetic susceptibility data in Figs.6 and 7. The two $B_{1g}$ Raman peaks also appear almost at the same temperature. Therefore, although the IP and OP are electronically independent of each other, they share the same bulk $T_c$ that is expected to be an average of the two $T_c$ values for the IP and OP. The idea of averaging $T_c$ is supported by the fact that $T_c$ does not decrease even when the system goes into the overdoped regime[27]. (See Fig.5.) Here, as the holes are doped over the optimum level in average, $T_c$ increases in the underdoped IP but decreases in the overdoped OP. As a result, the average $T_c$ remains constant. If the higher $T_c$ layer governs the bulk $T_c$, as expected in the case of inhomogeneous superconductors, the constant $T_c$ with doping in Fig.5 cannot be explained.

In summary, the presence of the underdoped IP together with the overdoped OP is unique to multilayer cuprates and it characterizes their physical properties. Through the interlayer coupling, the IP with the pseudogap could enhance a pair-breaking energy on the one hand, but it may suppress $T_c$, giving a large gap ratio (PE/$T_{c,max}$) on the other hand.

## 7. Conclusion

The characteristic features of multilayer cuprates were reviewed with a focus on $T_c$ and gap energy. There were several important points. (i) Each layer is independent, namely, the electronic bands of IP and OP are not hybridized, which is different from a band picture. This is due to the strong electron correlation in this system, and is proved by the band splitting in ARPES. (ii) The hole concentrations of IP and OP are different, which was first found in NMR, and confirmed by ARPES and RSS. This is a result of (i), namely, the many body effect. (iii) Nevertheless, there is a coupling between IP and OP, which gives a single bulk $T_c$ and a common PE($p$) line for $B_{1g}$ Raman peaks. (iv) The Raman measurement has revealed that both the pair-breaking energy and the gap ratio (PE/$T_{c,max}$) are larger in triple layer cuprates than in single and double layer cuprates. The large gap ratio may be caused by the suppression of bulk $T_c$ by the proximity effect from the pseudo-gapped IP. The combination of underdoped IP and overdoped OP characterizes the multilayer cuprate system, giving a large pair-breaking energy scale.

Acknowledgement

The authors thank S. Ideta for providing us the raw ARPES data of Bi2223 for our Raman spectra calculation.

References

[1] Bednorz J G and Müller A 1986 Z. Phys. B **64** 189.
[2] Anderson P W and Zou Z 1988 Phys. Rev. Lett. **60** 132.
[3] Chakravarty S and Anderson P W 1994 Phys. Rev. Lett. **72** 3859.
[4] Tajima S, Schützmann J, Miyamoto S, Terasaki I, Sato Y and Hauff R 1997 Phys. Rev. B **55** 6051.
[5] Tranquada J, Sternlieb B J, Axe, J D, Nakamura Y and Uchida S, 1995 Nature **375** 561.

[6] For a review, 2012 Physica C (special issue) 481 edited by Uchida S, Tranquada J, Fink J, Kivelson S and Tajima S.
[7] For a review, Timusk T and Statt B, 1999 Rep. Prog. Phys. **62** 61.
[8] Wang Y, Li L and Ong N P 2006 Phys. Rev. B **73** 024510.
[9] Li L, Wang Y, Komiya S, Ono S, Ando Y, Gu G D and Ong N P 2010 Phys. Rev. B **81** 054510.
[10] Dubroka A *et al.* 2011 Phys. Rev. Lett. **106** 047006.
[11] Uykur E, Tanaka K, Masui T, Miyasaka S and Tajima S 2014 Phys. Rev. Lett. **112** 127003, and references therein.
[12] Iyo A *et al.* 2007 J. Phys. Soc. Jpn. **76** 094711.
[13] Eisaki H *et al.* 2004 Phys. Rev. B **69** 064512.
[14] Fujita K *et al.* 2006 Phys. Rev. Lett **96** 097006.
[15] Wheatley J M, Hsu T C and Anderson P W 1988 Nature (London) **333** 121.
[16] Chakravarty S, Kee H and Volker K 2004 Nature (London) **428** 53.
[17] Kivelson S A 2002 Physica B **318** 61.
[18] For a review, Cooper S L and Grey K E 1994 in *Physical Properties of the High Temperature Superconductors* IV edited by Ginsberg D M (World Scientific, Singapore, 1994) 61.
[19] Kleiner R, Steinmeyer F, Kunkel G and Mueller P, 1992 Phys. Rev. Lett. 68 2394.
[20] Terasaki I, Sayo Y, Miyamoto S, Tajima S and Koshizuka N 1995 Phys. Rev. B **52** 16246.
[21] Mukuda H, Shimuzu S, Iyo A and Kitaoka Y 2012 J. Phys. Soc. Jpn. **81** 011008.
[22] Presland M R, Tallon J L, Buckley R G, Liu R S and Flower N E 1991 Physica C **176** 95.
[23] Tallon J L, Bernhard C, Shaked H, Hitterman R L and Jorgensen J D 1995 Phys. Rev. B **51** 12911.
[24] Iwai S, Mukuda H, Shimizu S, Kitaoka Y, Ishida S, Iyo A, Eisaki H and Uchida S 2014 JPS Conf. Proc. **1** 012105.
[25] Kwok W N *et al.* 1990 Phys. Rev. B **42** 8686.
[26] Fujii T, Watanabe T and Matsuda A 2001 J. Cryst. Growth **223** 175.
[27] Fujii T, Terasaki I, Watanabe T and Matsuda A 2002 Phys. Rev. B **66** 024507.
[28] Ideta S *et al.* 2010 Phys. Rev. Lett. **104** 227001.
[29] Vincini G, Tanaka K, Adachi T, Sobirey L, Miyasaka S, Tajima S, Adachi S, Sasalo N and Watanabe T 2019 Phys. Rev. B **98** 144503.
[30] Feng D L *et al.* 2001 Phys. Rev. Lett. **86** 5550.
[31] Tanaka K *et al.* 2006 Science **314** 5907.

[32] Lee W S *et al.* 2007 Nature **450** 81.
[33] Yoshida T *et al.* 2009 Phys. Rev. Lett. **103** 037004.
[34] Devereaux T P and Hackl R 2007 Rev. Mod. Phys. **79** 175.
[35] Tadatomo H, Masui T and Tajima S 2010 Phys. Rev. B **82** 224503.
[36] Although a part of the OvD109 sample seems to contain a lower $T_c$ phase, we think that it is due to the imperfect oxygenation of the inner part of crystal. It is expected that the crystal surface is well oxygenated, namely, well overdoped, which is probed by Raman scattering.
[37] Devereaux T P, Einzel D, Stadtlober B, Hackl R, Leach D H and Neumeier J J 1994 Phys. Rev. Lett. **72** 396.
[38] NygeunTrung H 2014 Doctor Thesis (Osaka University); NygeunTrung H, Tanaka K, Masui T, Miyasaka S, Tajima S and Sasagawa T 2013 Phys. Procedia **45** 41.
[39] Tanaka K, NygeunTrung H, Vincini G, Masui T, Miyasaka S, Tajima S and Sasagawa T 2019 J. Phys. Soc. Jpn. **88** 044710.
[40] Devereaux T private communication.
[41] Vincini G 2018 Doctor Thesis (Osaka University).
[42] Sacuto A, Gallais Y, Cazayous M, Messon M –A, Gu G D and Colson D 2013 Rep. Prog. Phys. **76** 022502.
[43] Sakai S et al. 2013 Phys. Rev. Lett. **111** 2.
[44] Tacon M Le, Scauto A, Georges A, Kotliar G, Gallais Y, Colson D and Forget A 2006 Nat. Phys. **2** 537.
[45] Oda M, Dipasupli R M, Momono N and Ido M 2000 J. Phys. Soc. Jpn. **69** 983.
[46] Kondo T, Khasanov R, Takeuchi T, Schmalian J and Kaminski A 2009 Nature **457** 296
[47] Loret B, Sakai S, Gallais Y, Cazayous M, Messon M A, Forget A, Colson D, Civelli M and Sacuto A 2016 Phys. Rev. Lett. **116** 197001.
[48] Schützmann J *et al.* 1997 Phys. Rev. B **55** 11118.
[49] Pavarini E, Dasgupta I, Saha-Dasgupta T, Jesen O and Andersen O K 2001 Phys. Rev. Lett. 87 047003.

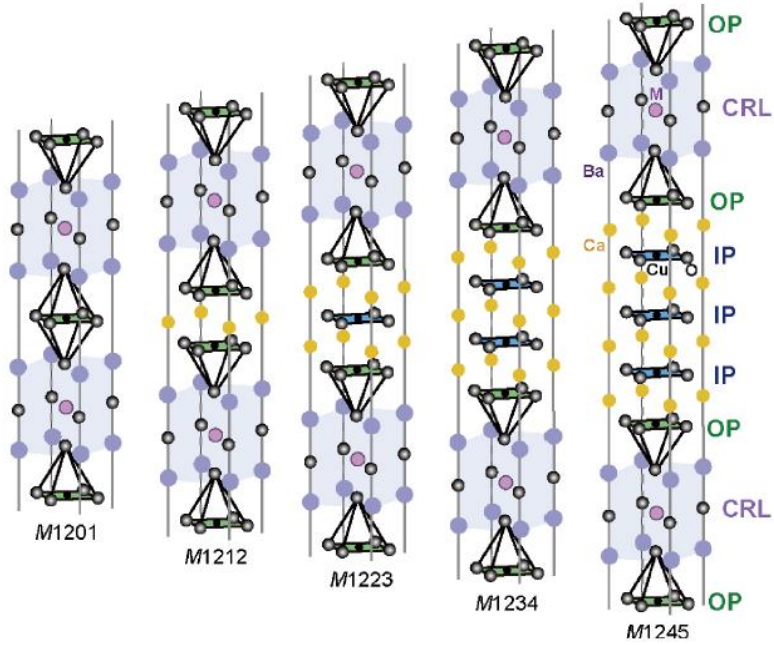

Figure 1. Crystal structure of $MBa_2Ca_{n-1}Cu_nO_{2n+2+\delta}$, M12(n-1)n cuprates, where M=Hg, Tl, and Cu. [21].   In a unit cell, outer $CuO_2$ planes (OP) in a five-fold pyramid coordination and inner $CuO_2$ planes (IP) in a four-fold square coordination are sandwitched by charge reservoir layers (CRL).

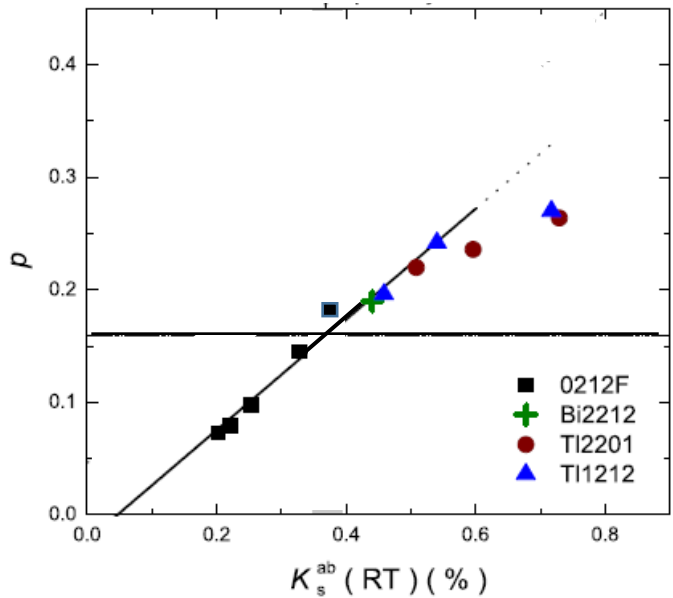

Figure 2. Relation between the carrier doping level $p$ and spin part of Knight shift with B//ab for several single and double layer cuprates. 0212F is $Ba_2Ca_{n-1}Cu_nO_{2n}(F_{1-y}O_y)_2$ with n=2. [21].

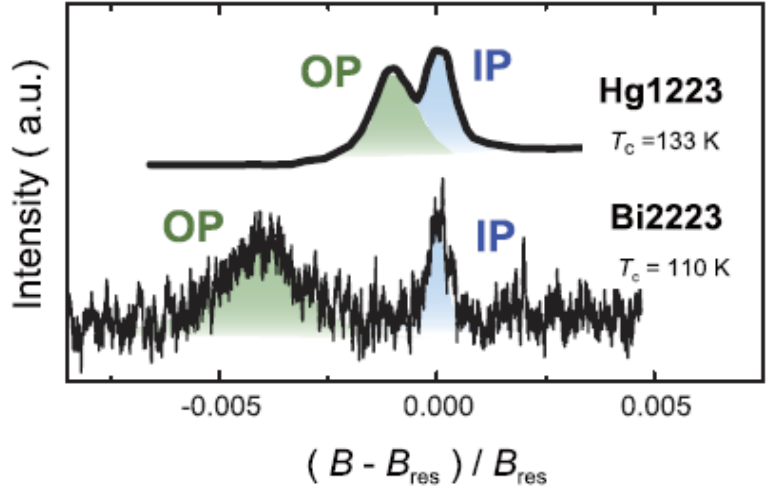

Figure 3. $^{63}$Cu-NMR spectra of Bi2223 single crystal and Hg1223 polycrystal. Here, a resonance field $B_{res}$ for $^{63}$Cu at B⊥c is 15.55T for Bi2223 and 10.95T for Hg1223. [24].

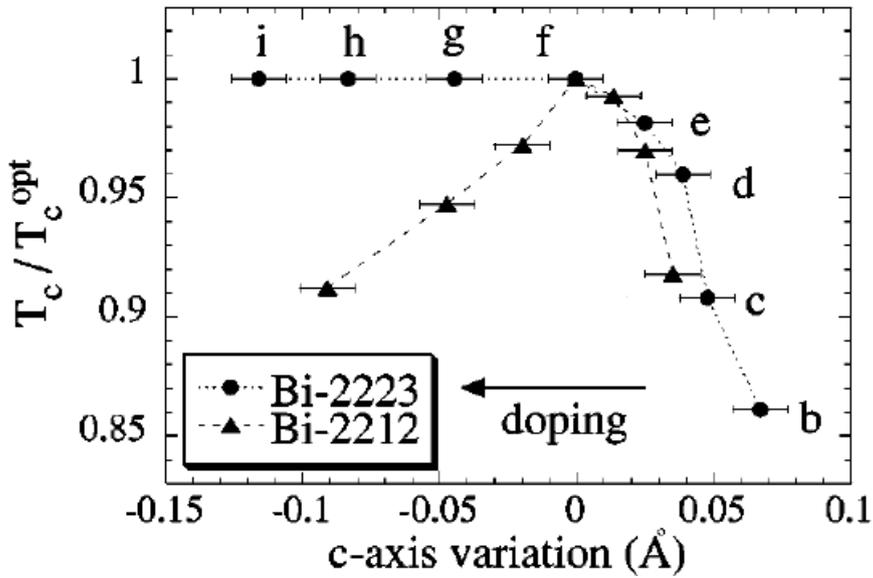

Figure 4. $T_c$ variation as a function of c-axis lattice parameter variation in Bi2223 and Bi2212. $T_c$ is normalized by the optimum value of each system. The notations of b to i are the sample names of Bi2223. [27].

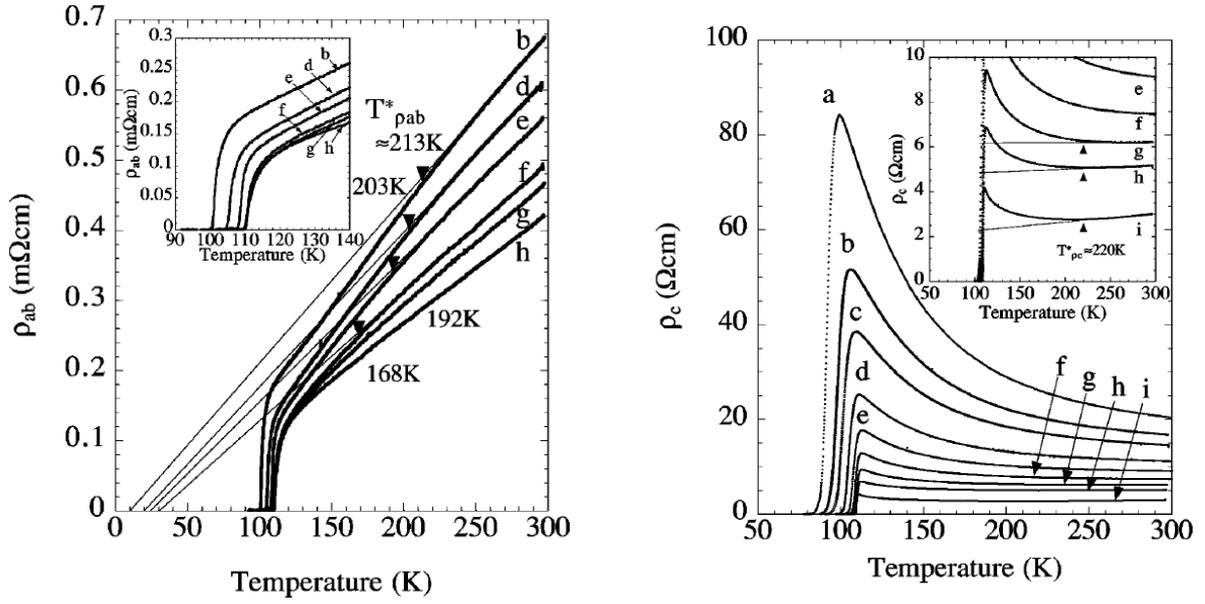

Figure 5. Temperature dependence of in-plane (left panel) and out-of-plane (right panel) electrical resistivity of Bi2223 single crystals with various oxygen contents. The sample notations (b-i) correspond to those in Fig.4. [27].

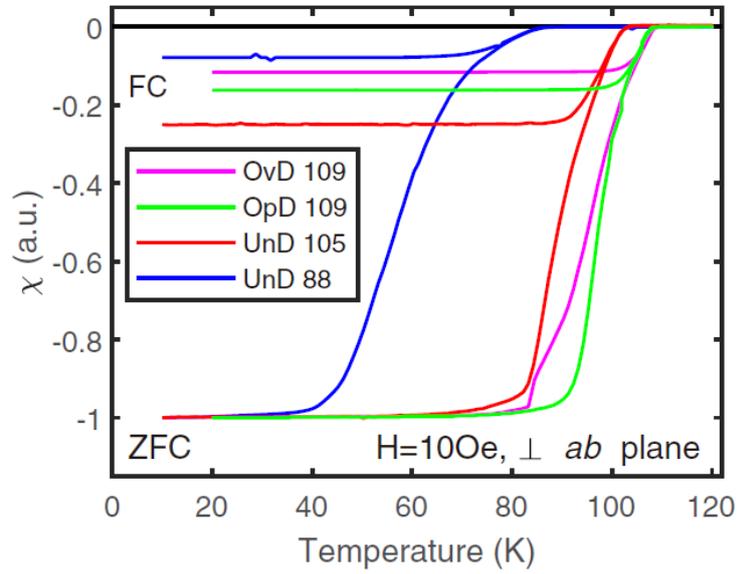

Figure 6. Magnetic susceptibility of Bi2223 with four doping levels. [29]

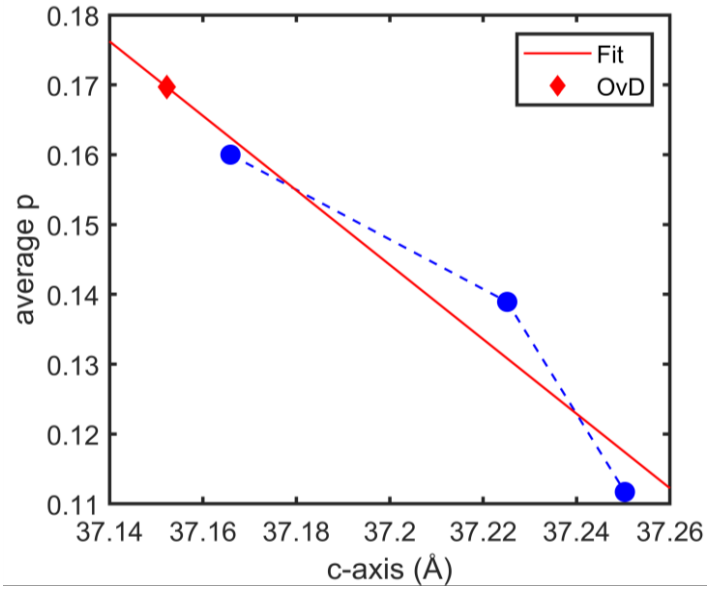

Figure 7. Average doping level $p$ versus $c$-axis lattice parameter. The blue circles indicate the data for the optimally doped sample with $T_c$=109K (OpD109) and the two underdoped samples with $T_c$ =105K (UnD105) and 88K (UnD88). The straight line is a fit to these three data points (the least square fit). From the $c$-axis lattice parameter determined by X-ray diffraction, the data point for the overdoped sample with $T_c$ =109K (OvD109) is expected as indicated by diamond symbol.[41]

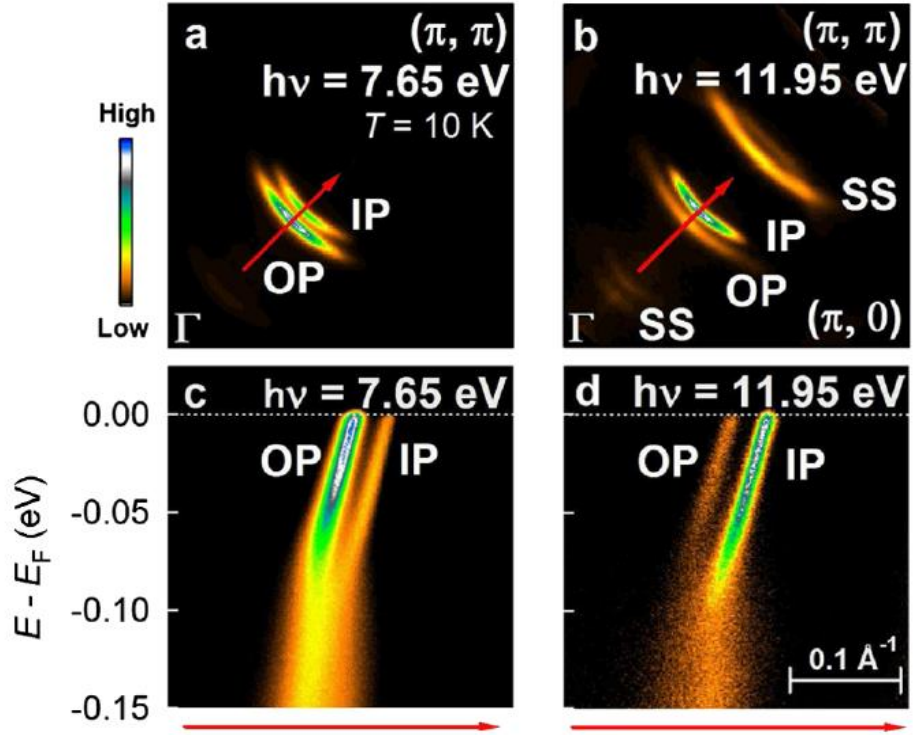

Figure 8. (a) and (b) Intensity plots of ARPES spectra with incident photon energy $h\nu$=7.65 eV and 11.95 eV in momentum space of Bi2223. (c) and (d) Band dispersions in the nodal direction indicated by arrows. When $h\nu$ is 7.65 eV (11.95 eV), the intensity of OP (IP) is enhanced. SS is a superstructure due to the BiO-layer modulation.[28].

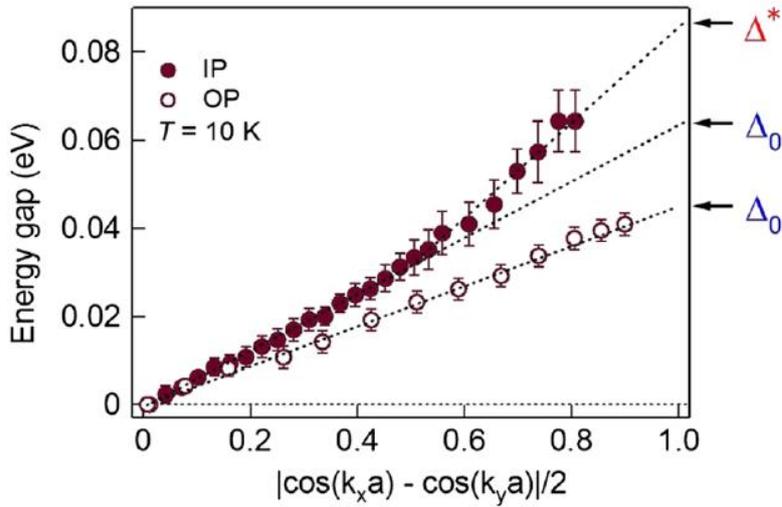

Figure 9. Momentum dependence of energy gaps for Bi2223. In the OP, the gap has a $d$-wave form, $\Delta(k) = \frac{1}{2}\Delta_0(cos k_x a - cos k_y a)$, while it is strongly enhanced towards $\Delta^*$ near the anti-nodal region in the IP. [28]

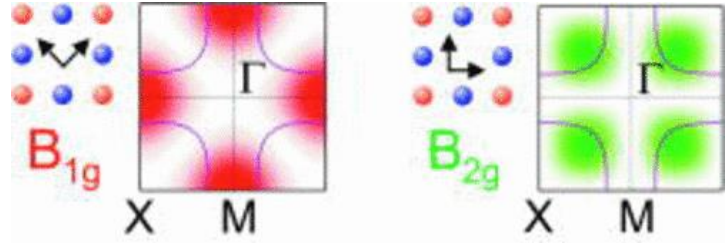

Figure 10. Schematic weighting of light-scattering transition for $B_{1g}$ and $B_{2g}$ polarizations in Raman scattering measurement.

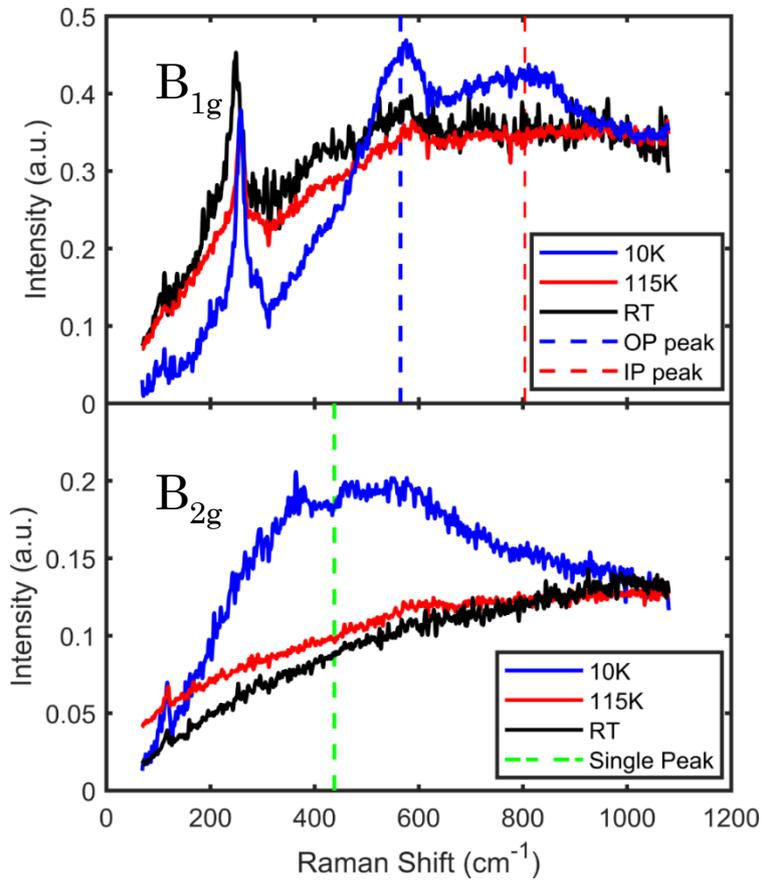

Figure 11. Raman scattering spectra of the OpD109 sample at room temperature (RT), 115K (just above $T_c$) and 10K (<< $T_c$) with $B_{1g}$ and $B_{2g}$ symmetry. The dashed lines indicate the peak positions. [41]

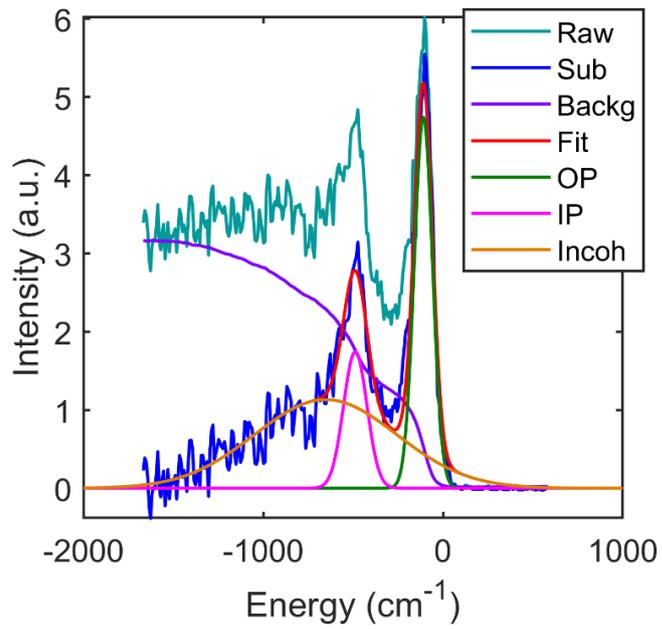

Figure 12. Extraction of IP and OP intensity from the raw ARPES data through fitting. From the raw data (Raw), the background (Backg) was subtracted. Then, the obtained intensity (Sub) was decomposed into three components (IP, OP and incoherent intensity). [41]

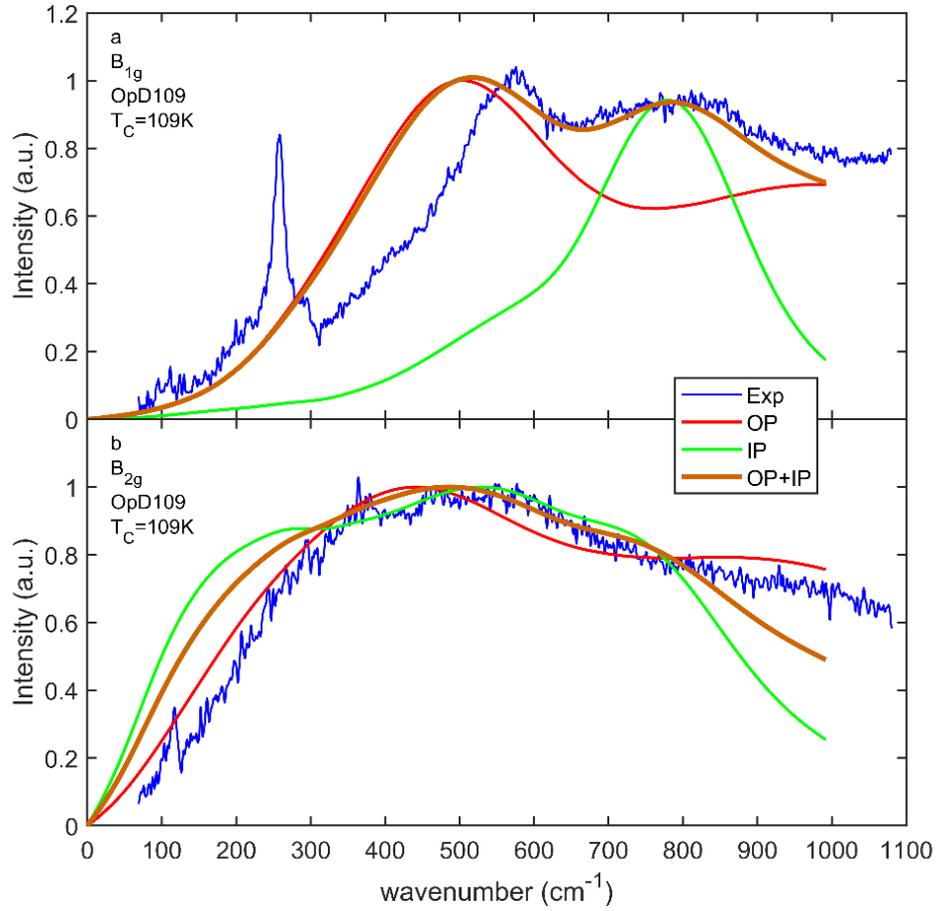

Figure 13. Comparison of the Raman spectra calculated from the ARPES (orange cruve) and the experimentally observed Raman spectra (blue curve) for $B_{1g}$ (a) and $B_{2g}$ (b) polarizations. The red and green curves are obtained from the ARPES data for the OP and IP, respectively. In the total curve IP+OP (thick orange), the contributions both from IP and OP are taken into account. [41]

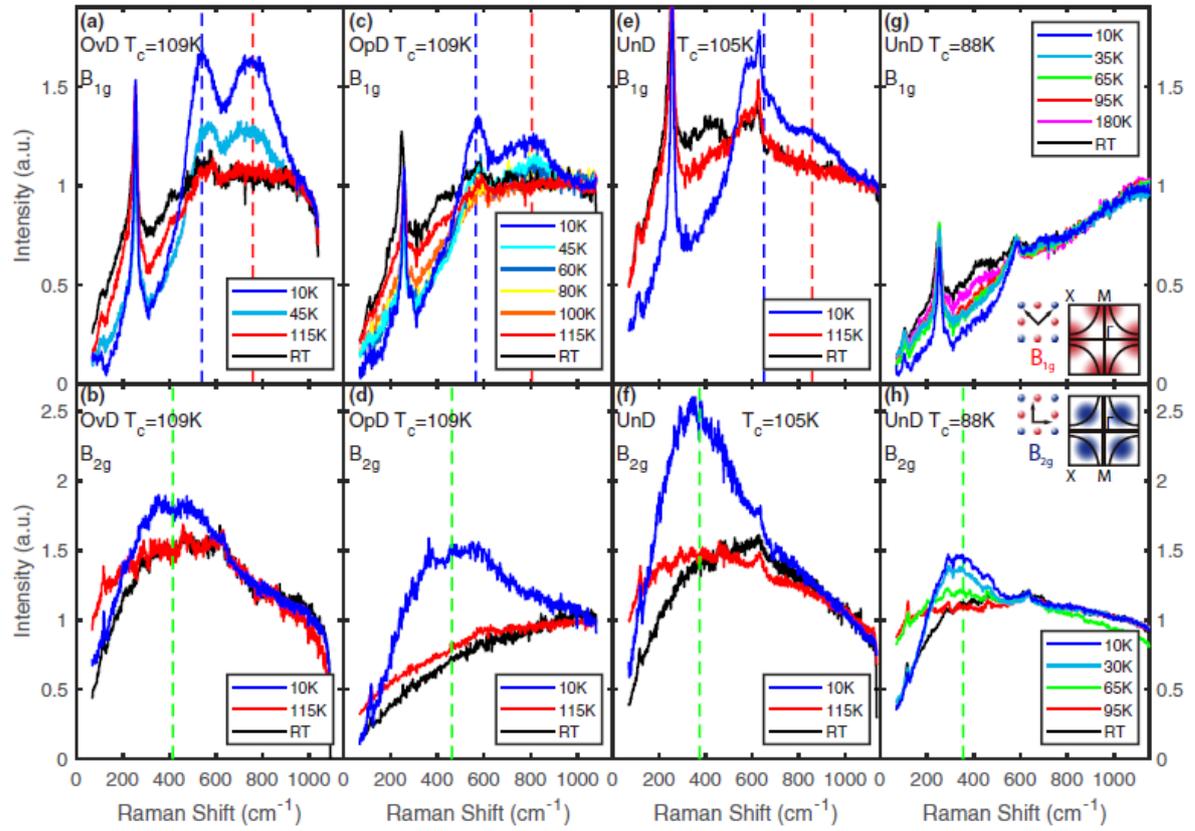

Figure 14. $B_{1g}$ and $B_{2g}$ Raman scattering spectra of OvD109 (a,b), OpD109 (c,d), UnD105 (e,f) and UnD88 (g,h) samples at several temperatures. The dashed lines indicate the peak positions. [29]

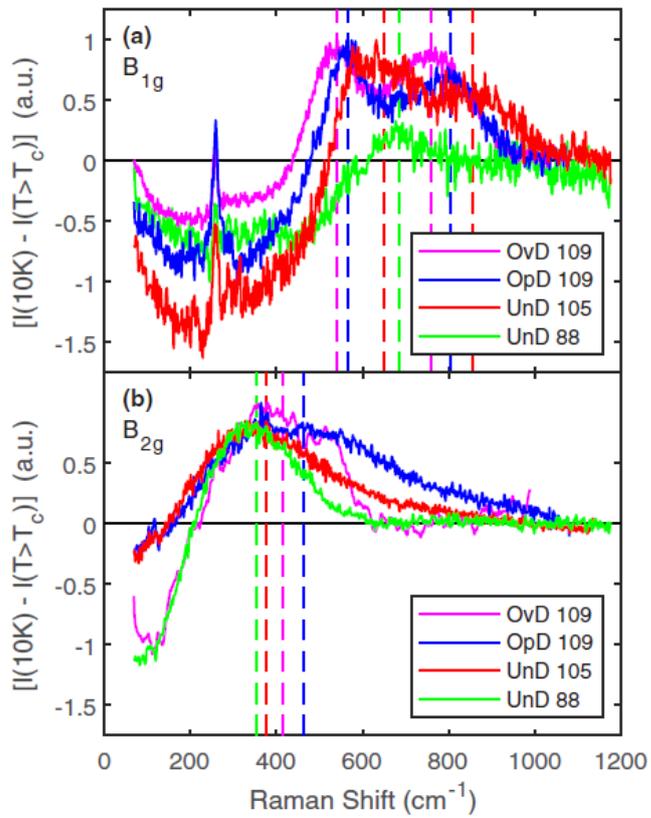

Figure 15. Difference spectra I(10K)-I($T>T_c$) with $B_{1g}$ and $B_{2g}$ polarizations. The $B_{1g}$ peak is visible also in UnD88 sample, although it is not clear in Fig.14. The dashed lines indicate the peak positions. [29]

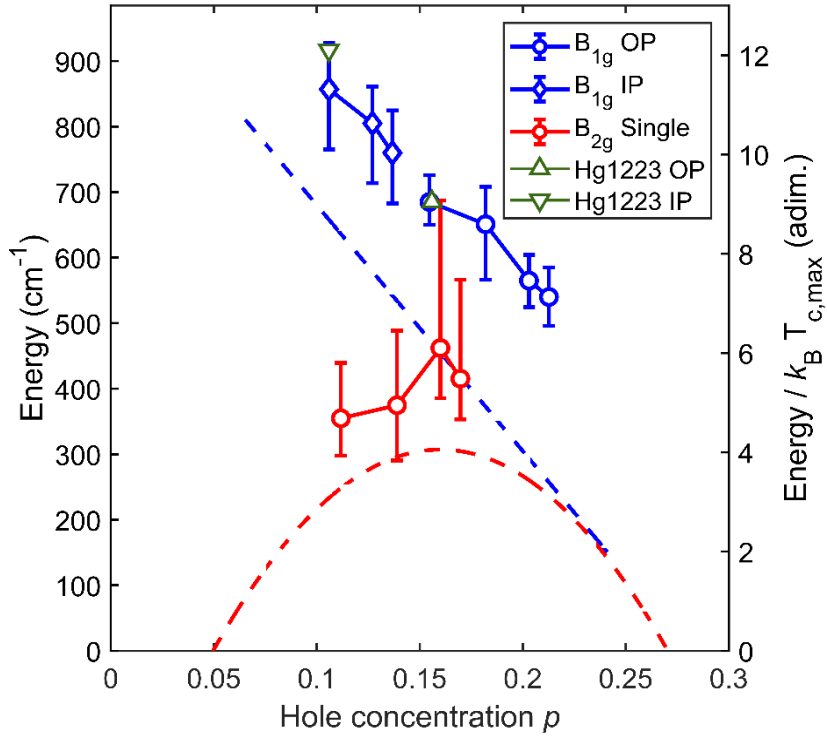

Figure 16. Doping dependence of the $B_{1g}$ and $B_{2g}$ pair-breaking peak energies (PE) and those normalized by the maximum $T_c$, PE/$k_B T_{c,\mathrm{max}}$. The data of PE/$k_B T_{c,\mathrm{max}}$ for Hg1223 $B_{1g}$ peak [47] are indicated by triangles, and those for the double layer cuprates are indicated by a dashed curve and line [44]. Note that $B_{1g}$ data are plotted as a function of $p$ of each layer, while $B_{2g}$ data are plotted as $p_{\mathrm{av}}$.

Table I  $T_c$, $c$-axis lattice parameter, average doping level ($p_{av}$), doping level ($p$) of OP and IP for the Bi2223 crystals used in ARPES (Tokyo crystal)[28] and Raman scattering experiments (OvD109, OpD109, UnD105, and UnD88) [41].

| Sample Name | $T_c$ (K) | $c$-axis (Å) | $p_{av}$ | $p$(OP) | $p$(IP) |
|---|---|---|---|---|---|
| Tokyo Crystal | 108 | – | 0.16 | – | – |
| OvD109 | 109 | 37.152 | 0.17 | 0.213 | 0.137 |
| OpD109 | 109 | 37.166 | 0.16 | 0.203 | 0.127 |
| UnD105 | 105 | 37.225 | 0.14 | 0.182 | 0.106 |
| UnD88 | 88 | 37.250 | 0.11 | 0.155 | 0.079 |